\documentclass[prd,superscriptaddress,%
nofootinbib,showpacs,preprintnumbers,floatfix]{revtex4}
\usepackage{epsfig}

\newcommand{\footfrac}[2]%
{\mbox{\footnotesize ${\displaystyle \frac{#1}{#2}}$}}
\begin{document}

\title{Target Mass Corrections Revisited}

\author{F. M. Steffens}

\address{   NFC - FCBEE - Universidade Presbiteriana Mackenzie,
            Rua da Consola\c{c}\~ao 930, 01302-907,
            S\~ao Paulo, SP, Brazil \\
            IFT - UNESP,
            Rua Pamplona 145, 01405-900,
            S\~ao Paulo, SP, Brazil}

\author{W. Melnitchouk}

\address{Jefferson Lab,
    12000 Jefferson Avenue,
    Newport News, VA 23606, USA}

\begin{abstract}
We propose a new implementation of target mass corrections to nucleon
structure functions which, unlike existing treatments, has the
correct kinematic threshold behavior at finite $Q^2$ in the $x \to
1$ limit. We illustrate the differences between the new approach
and existing prescriptions by considering specific examples for
the $F_2$ and $F_L$ structure functions, and discuss the broader
implications of our results, which call into question the notion
of universal parton distribution at finite $Q^2$.
\end{abstract}

\maketitle

\section{Introduction}

Deep inelastic lepton scattering is one of the most developed
tools with which to probe the quark and gluon structure of
hadrons. The theoretical framework within which the experiments
are analyzed is the operator product expansion (OPE). This is well
established in the Bjorken limit, which is defined by the
four-momentum transfer squared $Q^2$ and energy transfer $\nu$
being asymptotically large, with the ratio $x = Q^2/2M\nu$ (the
Bjorken scaling variable) fixed, and $M$ is the target mass.
Within this framework, global analyses of deep inelastic
scattering (DIS) and other experiments have allowed a vast array
of data to be described in terms of a universal, process
independent set of quark and gluon (or parton) distributions.

On the other hand, there exists a large body of data at lower
energies, at $Q^2 \lesssim 1-2$~GeV$^2$ (see {\em e.g.}
Ref.~\cite{PRep}), where the use of the asymptotic, Bjorken limit
formalism may be more questionable.
In addition to perturbative QCD effects generated by gluon radiation,
at low $Q^2$ effects arising from $1/Q^2$ power corrections become
increasingly important.
These are typically generated by multi-parton correlations,
and within the twist expansion are associated with higher twists
(in the OPE the twist of an operator is defined as its dimension
minus its spin).
Since they characterize the long-range nonperturbative
interactions between quarks and gluons, the higher twists contain
information on confinement dynamics, and are as such of intrinsic
interest to study.
Several recent analyses of structure function data at low $Q^2$
have extracted matrix elements of the higher twist operators
\cite{HTanal}.

Before one can reliably extract information on the higher twist
contributions, it is important to remove from the data corrections
arising from purely kinematic effects associated with finite
values of $Q^2/\nu^2 = 4 M^2 x^2/Q^2$.
These so-called ``target mass corrections'' (TMCs) are formally
related to twist-two operators, and hence contain no additional
information on nonperturbative multi-parton correlations.
Indeed, these have long been considered uninteresting and
believed to be well understood.
In the literature there are well known prescriptions for how to
remove the TMC corrections \cite{GP76,Bluemlein}.

The TMCs were first considered by Nachtmann \cite{Nac73}, who
showed that one could arrange the OPE so as to ensure that at a
given order in $1/Q^2$ only operators of a given twist would
appear.
At finite $Q^2$ the natural scaling variable, defined as the
fraction of the nucleon's light-front momentum ($p$) carried by
the parton ($k$), is then given by the Nachtmann scaling variable:
\begin{equation}
\xi (x, Q^2)\ \equiv\ { k^0 + k^z \over p^0 + p^z }\
     =\ { 2 x \over 1 + \sqrt{ 1 + 4 M^2 x^2/Q^2 } }\ .
\end{equation}
In the Nachtmann approach, one generalizes the Cornwall-Norton (CN)
moments of structure functions, derived in the Bjorken limit, to
finite $Q^2$.
A particular feature of these Nachtmann moments is that they are
supposed to factor out the target mass dependence of the structure
functions in a way such that its moments would equal the moments of
the corresponding parton distributions.

Later Georgi and Politzer (GP) \cite{GP76} calculated the CN moments
of the structure functions, taking into account the trace terms which
appear in the matrix elements of the twist-two operators, but which are
usually neglected in high $Q^2$ data analyses.
While the leading, twist-two piece of the structure function which
enters at ${\cal O}(1)$ is related to matrix elements of the quark
bilinear $\bar\psi \gamma^\mu \psi$, for example, the TMCs arise from
insertions of derivatives,
$\bar\psi \gamma^\mu D^{\gamma_{\mu_1}} \cdots D^{\gamma_{\mu_n}} \psi$,
which does not alter the twist.
This procedure generates a series in $M^2/Q^2$ when calculating
the CN moments.
Inverting these moments using the inverse Mellin transform, one 
arrives at a structure function depending on both $x$ and $M^2/Q^2$
\cite{GP76}.

Problems with the GP implementation of TMCs were soon identified,
however, by a number of authors
\cite{Ell76,GTW77,BJT79,JT80,Fra80}, in particular the so-called
``threshold problem''.
This pertains to the fact that if the parton distribution function
is a scaling function of $\xi$, then since the maximum kinematic
value of $\xi$ at any finite $Q^2$ is $\xi_0 \equiv \xi(x=1) < 1$,
the parton distribution is not defined in the unphysical region
between the elastic limit $\xi = \xi_0$ and $\xi = 1$.
De~Rujula {\em et al.} \cite{DGP77} argued that the problems can
be resolved by considering in addition higher twist operators.
They note that there is a nonuniformity in the limits as $n \to \infty$
and $Q^2 \to \infty$, and the appearance of higher twist effects
proportional to $n M^2/Q^2$ for the $n$-th moment signals the
breakdown of the entire approach at low $W$ ($\lesssim 2$~GeV).

Tung and collaborators \cite{BJT79,JT80} attempted to redress the
threshold problem by invoking an {\em ansatz} which smoothly
merges the perturbative QCD behavior of the moments at large $Q^2$
with the correct threshold behavior in the $n \to \infty$ limit.
As they note, however, such a prescription is not unique,
and in fact agrees with the standard OPE expansion only in the
$n \to \infty$ limit.

The proposed solution of De~Rujula {\em et al.} \cite{DGP77} to
the threshold problem implies that higher twist effects play an
important role at low $Q^2$. Recent experiments at Jefferson Lab
have shown, however, that the size of the higher twist effects is
actually quite small for the proton $F_2$ structure function, down
to relatively low $Q^2$ values ($Q^2 \sim 0.5-1$~GeV$^2$)
\cite{F2JL}.
The question which we address here is whether a self-consistent
formulation of TMCs can be made with {\em only} twist-two
contributions, without appealing to higher twist effects.
While not a proof, it seems plausible to us that, at least from
a purely theoretical perspective, it should be possible to obtain
an implementation of TMCs for a hypothetical case of negligible
higher twist effects, which would demand a consistent resolution
of the threshold problem for the twist-two part alone.
Such a view could be motivated by observing that even though
it is the same proton state that the twist-two and higher twist
terms originate from, in principle the matrix elements of the local
operators whose matrix elements characterize the twist expansion are
in fact independent.

While interesting in its own right, the question of how to
implement TMCs is also of practical importance, given the
high quality electron-nucleon structure function data at low and
moderate $Q^2$ which are being collected at Jefferson Lab \cite{F2JL}.
TMCs are also vital in analyzing neutrino scattering data
\cite{Kretzer}, much of which are taken at relatively low
energies, and must be understood if one is to extract reliable
information on neutrino oscillations for instance.
For spin-dependent scattering, TMCs have also been calculated for
the $g_1$ and $g_2$ structure functions \cite{Bluemlein},
and recently for spin-1 targets such as the deuteron \cite{Detmold}.

The pertinent question is whether the Nachtmann moment of the
twist-two part of the structure function is $Q^2$ independent,
as supposed in the original formulation \cite{Nac73}.
In Sec.~II we review the standard derivation and results for
TMCs within the operator product expansion.
We outline the problems associated with the standard approach,
and suggest an alternative formulation designed to avoid the
unphysical threshold problem.
Earlier work \cite{Fra80} did indeed find that the Nachtmann
moments do not account for all possible (leading twist) $M^2/Q^2$
effects.
However, we suggest a prescription where the $M^2/Q^2$ dependence
of the Nachtmann moments of the structure functions does equal,
to very high accuracy, the $M^2/Q^2$ dependence of the moments
of the quark distributions, for all $Q^2$.
Numerical results are presented in Sec.~III, where we compare the
$x$ dependence of the $F_2$ and $F_L$ structure functions using
the various TMC prescriptions, and examine the onset of scaling
in terms of the Nachtmann moments of the structure functions.
In Sec.~IV we summarize our findings and discuss the broader
implications of our results for the interpretation of parton
distributions at finite $Q^2$.

\section{Operator Product Expansion}

We begin this section by firstly reviewing the pioneering work on
target mass corrections as obtained by Georgi and Politzer \cite{GP76}.
We will consider the case of unpolarized scattering from a spin-1/2
nucleon, which is described by two structure functions, $F_1(x,Q^2)$
and $F_2(x,Q^2)$ (or alternatively $F_2$ and the longitudinal structure
function $F_L$).
We shall focus on the $F_2$ structure function, but later generalize
the discussion to include also $F_L$.

The two standard moments of structure functions encountered in the
literature are the Cornwall-Norton and Nachtmann moments.
The Cornwall-Norton moments of $F_2$ are given by:
\begin{eqnarray}
M_2^n(Q^2) &=& \int_0^1 dx\ x^{n-2}\ F_2(x,Q^2)\ , \label{e2}
\end{eqnarray}
and are appropriate for the region $Q^2 \gg M^2$.
The Nachtmann moments, on the other hand, take into account finite
$M^2/Q^2$ corrections to the Bjorken limit, and are given by:
\begin{eqnarray}
\mu_2^n(Q^2) &=& \int_0^1 dx\ { \xi^{n+1} \over x^3 }
    \left[ { 3 + 3(n+1)r + n(n+2)r^2 \over (n+2)(n+3) } \right]
    F_2(x,Q^2)\ ,
\label{e9}
\end{eqnarray}
with $r=\sqrt{1 + 4 x^2 M^2/Q^2}$. The essential difference between
the CN and Nachtmann moments comes from the trace terms appearing
in the matrix elements of operators of definite spin, which are
disregarded in the CN approach, but kept in the Nachtmann approach.
The Nachtmann moments are constructed such that from the infinite
operators of twist-2 and different spin contained in the trace terms,
only the operators of spin $n$ contribute for the $n-2$ moment of the
structure function.

The Nachtmann and CN moments can be related by expanding the moments
in powers of $1/Q^2$.
Expanding $\mu_2^n$ to ${\cal O}(1/Q^6)$, one has:
\begin{eqnarray}
\mu_2^n(Q^2)
&=& M_2^n(Q^2)\ -\ { n(n-1) \over n+2 } {M^2 \over Q^2} M_2^{n+2}(Q^2)\
 +\ {n(n^2 - 1) \over 2 (n+3)} {M^4 \over Q^4} M_2^{n+4}(Q^2)\
 -\ {n (n^2 - 1) \over 6} {M^6 \over Q^6} M_2^{n+6}\
 +\ \cdots
\end{eqnarray}
Note that there is a mixing between the lower and higher moments.
To this order we can also express the CN moments in terms of the
Nachtmann moments:
\begin{eqnarray}
M_2^n(Q^2)
&=& \mu_2^n(Q^2)\ +\ { n(n-1) \over n+2 } {M^2 \over Q^2} \mu_2^{n+2}(Q^2)\
 +\ {n (n^2 - 1) (n+2) \over 2(n+3) (n+4)} {M^4 \over Q^4} \mu_2^{n+4}(Q^2)
\nonumber \\*
&+& {n (n^2-1) (n+2) (n+3) \over 6 (n+5) (n+6)}
    {M^6 \over Q^6} \mu_2^{n+6}\
 +\ \cdots
\label{eq:M2expand}
\end{eqnarray}

In the work of GP, the moment of the leading twist part of the $F_2$
structure function, corrected for target mass effects, can be written
to order $1/Q^6$ as:
\begin{eqnarray}
M_n^{\rm GP}(Q^2)
&=& A_n\ +\ \frac{n(n-1)}{n+2}\frac{M^2}{Q^2}\ A_{n+2}\
 +\ {n (n^2 - 1) (n+2) \over 2(n+3) (n+4)} {M^4 \over Q^4} A_{n+4}(Q^2)
\nonumber \\*
&+& {n (n^2-1) (n+2) (n+3) \over 6(n+5)(n+6)} {M^6 \over Q^6} A_{n+6}\
 +\ \cdots\ ,
\label{eq:M2GP}
\end{eqnarray}
where $A_n$ is the $n$-th moment of a distribution function $F(y)$:
\begin{equation}
A_n = \int_0^{y_0} dy\ y^n\ F(y)\ .
\label{eq:An}
\end{equation}
Here the function $F(y)$ is related to the usual quark distribution
$q(y)$ by $q(y) \equiv y F(y)$, and the upper limit of integration
$y_0$ is the maximum value at which the quark distribution has physical
support.
Again in Eq.~(\ref{eq:M2GP}) there is a mixture between lower and
higher moments.
Comparing Eqs.~(\ref{eq:M2expand}) and (\ref{eq:M2GP}), one can show
that, at least to ${\cal O}(1/Q^6)$, the Nachtmann moments are
equivalent to the moments of the distribution $F(y)$:
\begin{equation}
\mu_2^n \equiv A_n\ . \label{eq:idtmc}
\end{equation}
This reflects the fact that the Nachtmann moments are constructed
to protect the moments of the structure functions from target mass
effects, thereby allowing them to be identified directly with the moments
of the quark distributions.

The $F_2$ structure function appearing in Eqs.~(\ref{e2}) and (\ref{e9})
must itself be corrected for target mass effects, and to this end we
will follow the procedure in GP \cite{GP76}, albeit with one exception.
While GP write the upper limit of integration in Eq.~(\ref{eq:An})
as $y_0 = 1$, we will define the upper limit of the integrals as
the maximum value allowed by kinematics, $y_0 = y(x=1)$.
Following GP, we can then rewrite Eq.~(\ref{eq:An}) as \cite{GP76}:
\begin{equation}
\frac{A_{n+2j}}{(n+2j)(n+2j-1)} = \int_0^{y_0} dy\ y^{n+2j-2}\ G(y)\ ,
\label{e11}
\end{equation}
with $G(y)$ given by:
\begin{equation}
G(y) = \int_y^{y_0 } dy^\prime\ H(y^\prime)
     = \int_y^{y_0}dy^\prime\
       \int_{y^\prime}^{y_0} dy^{\prime \prime}\ F(y^{\prime \prime})\ .
\end{equation}
This result follows from the fact that:
\begin{equation}
\int_0^{y_0} dy\ y^{n+2j-2}\ G(y)
= \frac{y^{n+2j-1}}{n+2j-1} G(y)\big|_0^{y_0}
- \int_0^{y_0} dy\ \frac{y^{n+2j-1}}{n+2j-1}
  \frac{\partial G(y)}{\partial y}\ ,
\label{e13}
\end{equation}
and because $G(0) = G(y_0) = 0$, one is left with the second term only.
Integrating the RHS of Eq.~(\ref{e13}) again, we recover Eq.~(\ref{e11}).
To obtain the $x$ (or $\xi$) dependence of the structure functions,
we can invert the moment as in GP, arriving at same results but with
a modified upper limit in the integrals.
Specifically, we recover for the $F_2$ structure function \cite{GP76}:
\begin{equation}
F_2(x,Q^2)\
=\ \frac{\xi^2 (1 - a^2 \xi^2)}{(1 + a^2 \xi^2)^3} F(\xi)\
+\ 6 a^2 \frac{\xi^3 (1 - a^2 \xi^2)}{(1 + a^2 \xi^2)^4} H(\xi)\
+\ 12 a^4 \frac{\xi^4 (1 - a^2 \xi^2)}{(1 + a^2 \xi^2)^5}G(\xi)\ ,
\label{eq:F2}
\end{equation}
where $a \equiv M/Q$.
The CN moments of the target mass corrected $F_2$ structure function are
then given by \cite{GP76}:
\begin{equation}
M_2^n (Q^2)\
=\ \int_0^1 dx x^{n-2}F_2(x,Q^2)\
=\ \sum_{j=0}^\infty
   \left(\frac{M^2}{Q^2}\right)^j \frac{(n+j)!}{j! (n-2)!}
   \frac{A_{n+2j}}{(n+2j)(n+2j-1)}\ .
\label{e15}
\end{equation}
To calculate the Nachtmann moments, we rewrite Eq.~(\ref{e9}) in terms of
$\xi$:
\begin{equation}
\mu_2^n(Q^2)\
=\ \int_0^{\xi_0} d\xi\ \xi^{n-2}\
   \frac{(1 + a^2 \xi^2)^3}{1 - a^2 \xi^2}\ F_2(\xi,Q^2)
   \left[1 - \frac{3(r - 1)}{r^2(n+2)} - \frac{3(r-1)^2}{r^2(n+3)}
   \right]\ ,
\end{equation}
with $F_2(\xi,Q^2)$ given by Eq.~(\ref{eq:F2}), and
\begin{equation}
\xi_0\ =\ \xi(x=1)\ =\ { 2 \over { 1 + \sqrt{1 + 4 M^2/Q^2} } }\ .
\end{equation}
In the following section we will examine the extent to which the
Nachtmann moments of $F_2$ correspond to the moments $A_n$ of the
quark distribution function, for different functional forms of $F(\xi)$,
and quantify the effect of the kinematic thresholds.
Central will be the interpretation of the function $F(\xi)$ itself.

Before proceeding, for completeness we also give the results for the
longitudinal structure function, $F_L$, and its moments.
In the $Q^2 \to \infty$ limit, $F_L = 0$, while at finite $Q^2$ the
TMCs render $F_L$ nonzero.
Of course, higher order perturbative QCD corrections which depend on
$\alpha_s$ also give rise to a nonzero $F_L$, as do higher twist effects.
However, we artificially set both of these to zero in order to isolate
the effects of TMCs on $F_L$ explicitly.
Following a similar procedure as for $F_2$ above, we can write the
longitudinal structure function as:
\begin{equation}
F_L(\xi,Q^2)\
=\ \frac{2 a^2 \xi^2}{(1 + a^2 \xi^2)^2}H(\xi)\
+\ \frac{4 a^4 \xi^3}{(1 + a^2 \xi^2)^3}G(\xi)\ ,
\end{equation}
which clearly vanishes as $a \to 0$ (or $Q^2 \to \infty$).
The corresponding Nachtmann moments are then given by:
\begin{equation}
\mu_L^n(Q^2)\
=\ \int_0^{\xi_0} d\xi\ \xi^{n-2} (1 - a^4 \xi^4)
   \left( F_L(\xi,Q^2)
    + \frac{4 a^2 \xi^2}{(1 - a^2 \xi^2)}
      \frac{(n+1)(1 - a^2 \xi^2) - 2(n+2)}{(n+2)(n+3)}\
      F_2(\xi,Q^2)
   \right)\ .
\end{equation}
Having derived analytic expressions for the $F_2$ and $F_L$ structure
functions and their moments, in the next section we present numerical
results using several different prescriptions for the $\xi$ dependence
of the quark distribution function.

\section{Target Mass Corrections}

The main purpose of this work is to analyze phenomenologically what is
the best procedure to incorporate TMCs in the analysis of structure
functions.
More specifically, we address the question of which procedure is most
effective in rendering the moments of the leading twist structure
functions equal to the moments of the quark distributions at finite
$Q^2$.
We consider whether there is any sizable difference for the moments when
the upper limit of the integrals in $G(y)$ and $H(y)$ is 1 or $\xi_0$.
In particular, since the twist-two part of the deep inelastic cross
section should be zero at $x=1 \; (\xi = \xi_0)$, we study the impact
of a vanishing parton distribution at $\xi_0$.

\subsection{Prescriptions}

To address these issues, in this section we present several
prescriptions for the implementation of target mass corrections,
and discuss their limitations and practical consequences.
We consider three scenarios:

(A)
Integrate a quark distribution:
\begin{eqnarray}
q(\xi) &=& {\cal N}\ \xi^{-1/2} (1-\xi)^3\ 
\end{eqnarray}
from 0 to 1 (specifically, in the integrals for $A_n$, $H(\xi)$ and
$G(\xi)$).
Here the normalization ${\cal N}$ ensures that the distribution
integrates to unity.
We denote this prescription the ``standard TMC'' (sTMC).

(B)
Integrate a modified distribution which vanishes for $\xi > \xi_0$,
as implied by Eq.~(\ref{eq:An})%
    \footnote{We believe this was also the implication
    of De~R\'{u}jula {\em et al.} \cite{DGP77}}:
\begin{eqnarray}
q(\xi) &=& {\cal N}\ \xi^{-1/2} (1-\xi)^3\ \Theta(\xi - \xi_0)\ .
\end{eqnarray}
We denote this prescription the ``modified TMC'' (mTMC).

(C)
Use a ``threshold dependent'' (TD) quark distribution which
vanishes in the physical limit:
\begin{eqnarray}
q^{\rm TD}(\xi) &=& {\cal N}\ \xi^{-1/2} (\xi_0 - \xi)^3\ .
\end{eqnarray}

\begin{figure}[htb]
\epsfig{file=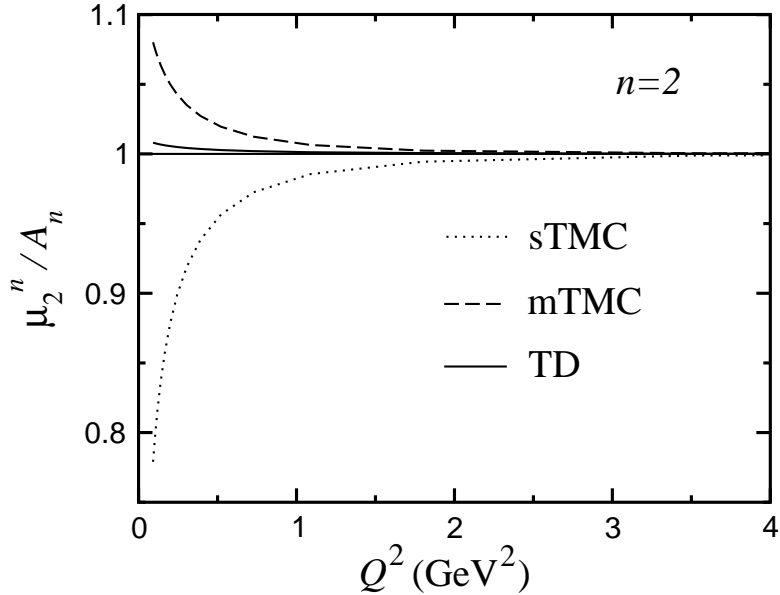,height=12cm,angle=-90}
\caption{Ratio of the $n=2$ Nachtmann moment of the $F_2$ structure
    function and the $n=2$ moment of the quark distribution,
    as a function of $Q^2$.
    The curves correspond to prescriptions A [``sTMC''] (dotted),
    B [``mTMC''] (dashed) and C [``TD''] (solid).}
\label{fig:An2}
\end{figure}

Note that because of the upper limit in Eq.~(\ref{eq:An}), $A_n$
itself will be $M^2/Q^2$ dependent for prescriptions B and C.
The results for the ratio $\mu_2^n/A_n$ of the $n=2$ moments
are displayed in Fig.~1 for the three cases, with prescriptions
A, B and C corresponding to the dotted, dashed and solid curves,
respectively.
Comparing the sTMC and mTMC results, one can see a reduced $Q^2$
dependence when the integrals are restricted to $\xi < \xi_0$.
However, a much more dramatic change occurs when the quark
distribution is constrained to vanish at $\xi_0$.
This renders the Nachtmann moment almost equal to the moment of
the quark distribution for virtually all $Q^2$ considered.
Certainly for $Q^2 > 1$~GeV$^2$ there is no visible deviation of
the ratio from unity.
Even for very small $Q^2$, $Q^2 \sim 0.3$~GeV$^2$, the ratio
differs from unity by only $\sim 0.7\%$ (of course the OPE itself
may not be valid at such low values of $Q^2$).

\begin{figure}[htb]
\epsfig{file=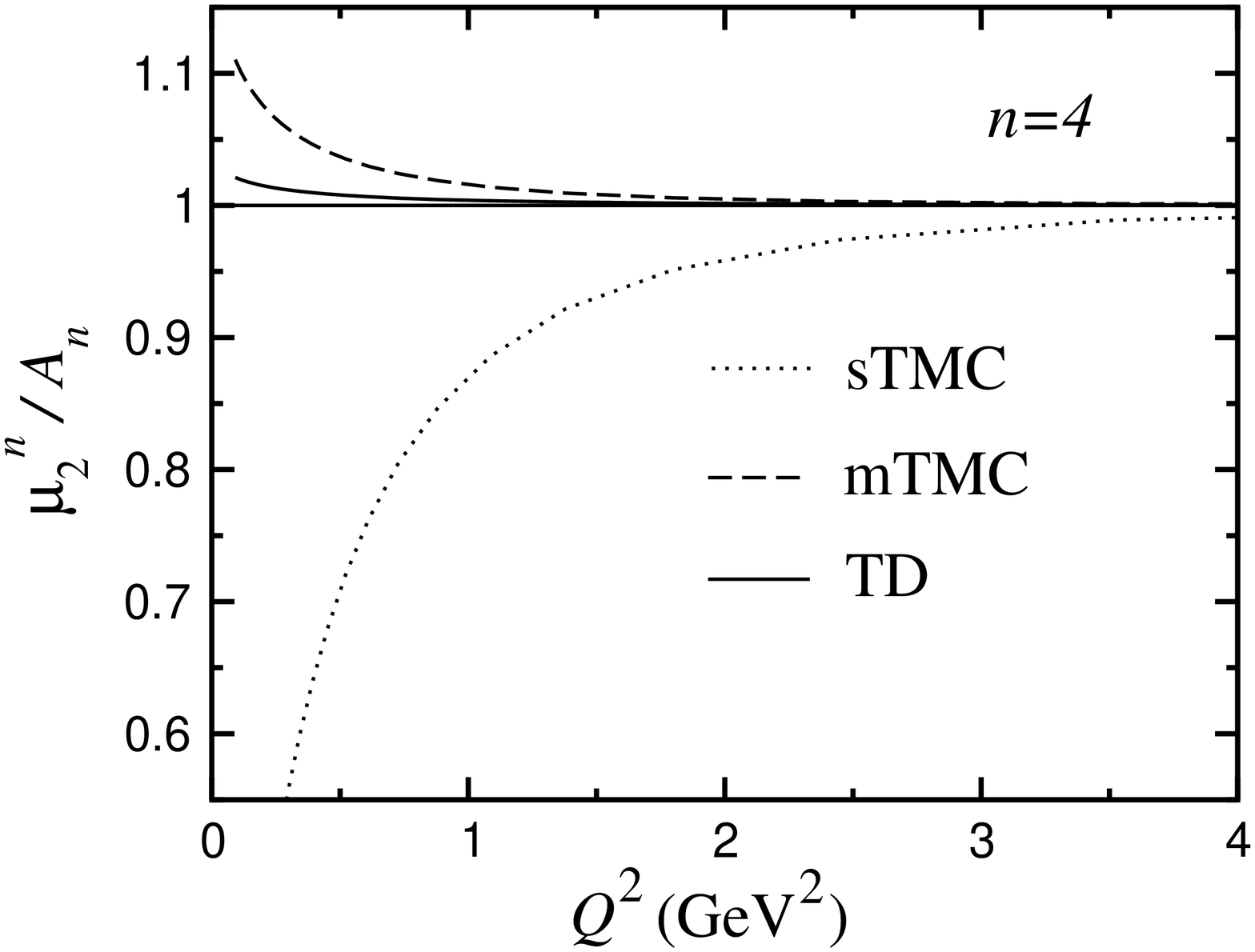,height=12cm,angle=-90}
\epsfig{file=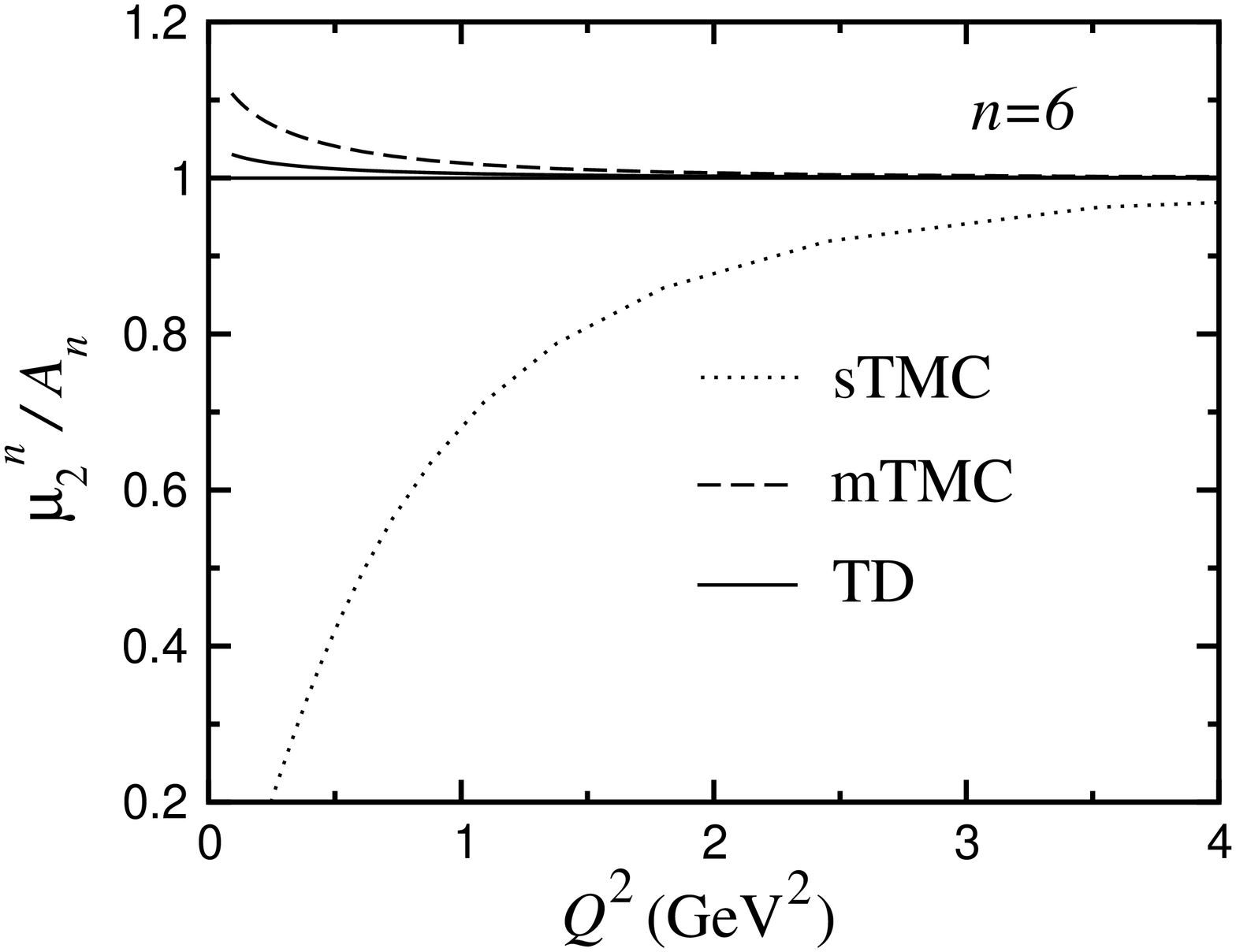,height=12cm,angle=-90}
\caption{Ratios of the $n=4$ (upper graph) and $n=6$ (lower graph)
    Nachtmann moment of the $F_2$ structure function and the
    corresponding moments of the quark distribution,
    as a function of $Q^2$.
    The curves are as in Fig.~\ref{fig:An2}.}
\label{fig:An4_6}
\end{figure}

Similarly, the ratios for the $n=4$ and $n=6$ moments are shown in
Fig.~\ref{fig:An4_6}.
The deviation of the ratio from unity for the sTMC approach is
between $10\% - 20\%$ for $Q^2 \lesssim 1$~GeV$^2$, while that for
the modified TMC with prescription B is of the order of 5\% over the
same $Q^2$ region.
On the other hand, for the threshold dependent prescription C, the
deviation from unity remains around 1\% even at these low $Q^2$ values.

A consequence of prescription C is that the moments of the parton
distribution are $Q^2$ dependent.
This seems to be an inevitable consequence if the Nachtmann moments of
the structure function are to be equal to the moments of the parton
distribution for all $Q^2$.
Note that this $Q^2$ dependence is not of higher twist or perturbative
QCD origin, but arises solely from kinematics.
Nevertheless, this avoids the more serious problems which arise within
the sTMC approach (prescription A), where the Nachtmann moments below
$Q^2 \sim 1$~GeV$^2$ start to deviate significantly from the moments of
the quark distributions.
In addition, in the sTMC formulation one is faced with the so-called
``threshold problem''.
Namely, if the moments $A_n$ of the quark distributions are $Q^2$
independent, then one should have:
\begin{equation}
\int_0^1 d\xi\ \xi^n\ F(\xi,Q_1^2)
= \int_0^1 d\xi\ \xi^n\ F(\xi,Q_2^2)
\label{eq:anequality}
\end{equation}
for any two momentum scales $Q_1^2$ and $Q_2^2$.
Since $F(\xi,Q^2)$ must vanish in the kinematically forbidden region
$\xi > \xi_0$, the equality in Eq.~(\ref{eq:anequality}) implies that
the function must be zero for both $\xi > \xi_0(Q_1^2)$ and
$\xi > \xi_0(Q_2^2)$.
If $Q_1^2 < Q_2^2$, in which case $\xi_0(Q_1^2) < \xi_0(Q_2^2)$,
this implies that $F(\xi,Q_2^2)$ should vanish in the range
$\xi_0(Q_1^2) < \xi < \xi_0(Q_2^2)$.
However, there is no physical reason for it to do so here, and this
therefore leads to an unphysical constraint.

De~R\'ujula {\em et al.} \cite{DGP77} address this problem by pointing
out that higher twist contributions play an ever more important role at
low $Q^2$, and their neglect makes any leading twist analysis at large
$\xi$ incomplete.
On the other hand, the philosophy inherent in prescription C is that
the equality (\ref{eq:anequality}) should not be expected to hold,
simply because the $\xi$ dependence, and also the normalization,
of the quark distributions is $M^2/Q^2$ dependent:
\begin{equation}
\int_0^{\xi_0(Q_1^2)} d\xi\ \xi^n\ F(\xi,Q_1^2)\
\neq\
\int_0^{\xi_0(Q_2^2)} d\xi\ \xi^n\ F(\xi,Q_2^2)\ .
\label{aninequality}
\end{equation}
In the following section we will contrast the various prescriptions
by studying their effects on the structure functions numerically.

\subsection{Numerical Results for Structure Functions}

\begin{figure}[htb]
\epsfig{file=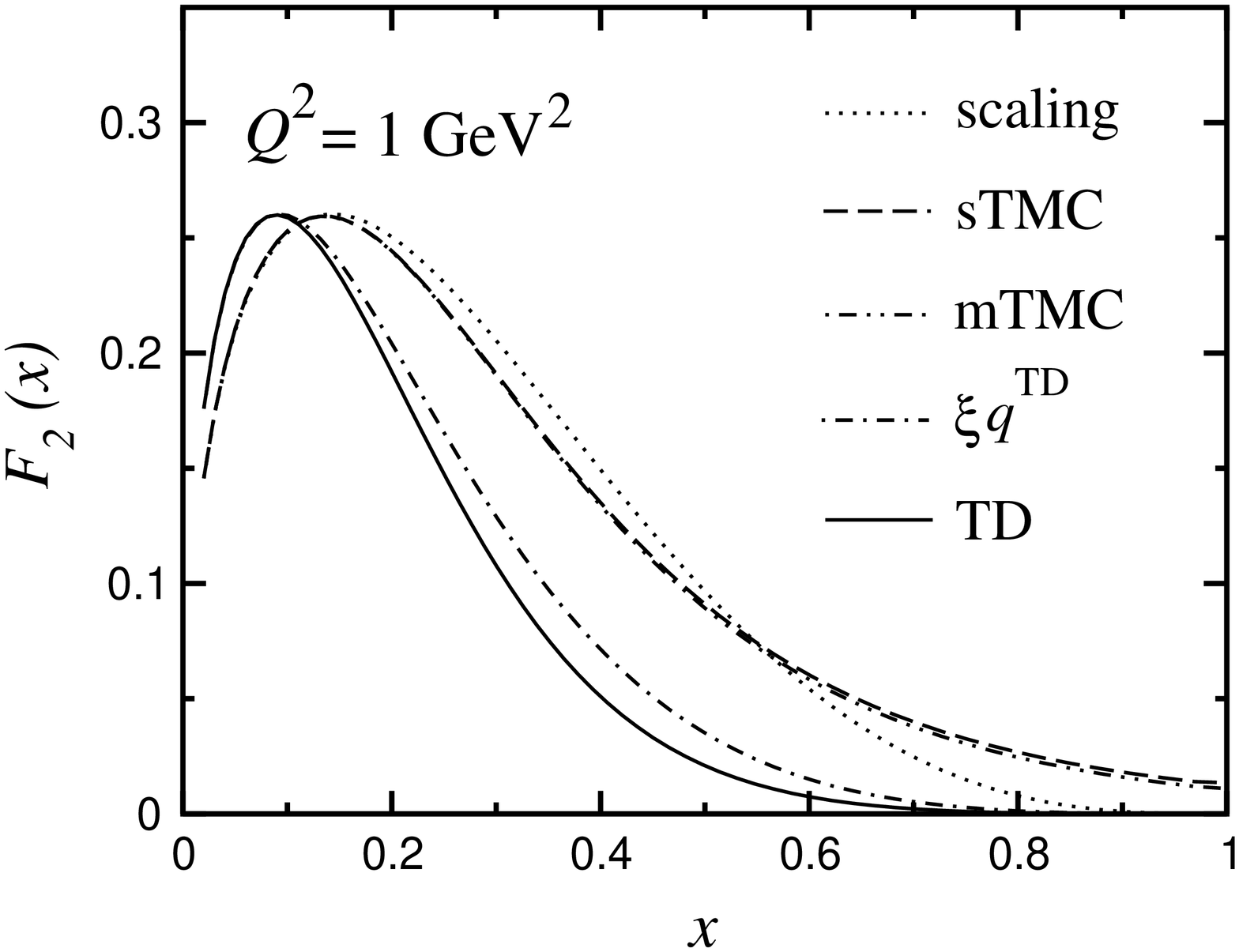,height=12cm,angle=270}
\epsfig{file=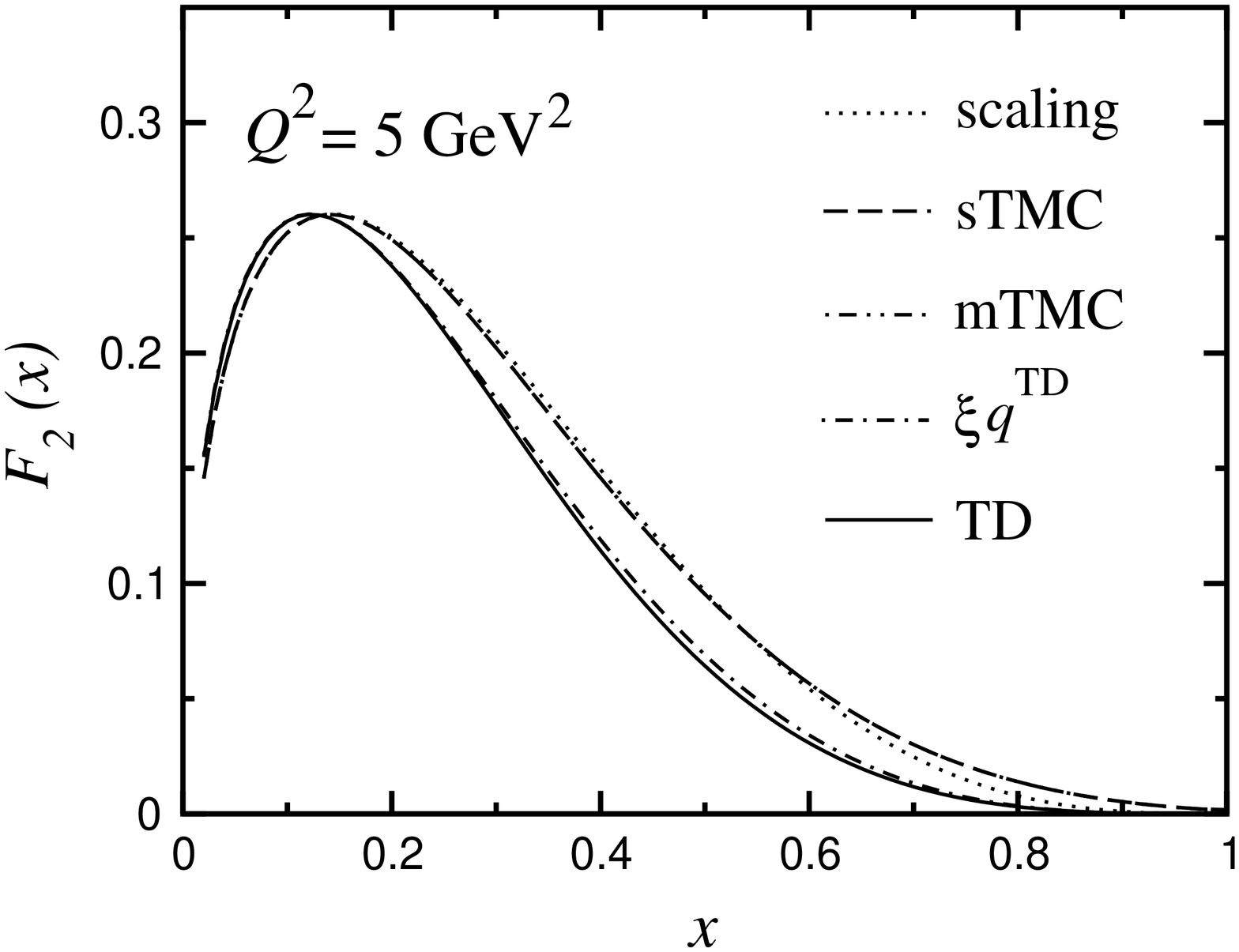,height=12cm,angle=270}
\caption{The $x$ dependence of the $F_2$ structure function
    at $Q^2=1$~GeV$^2$ (upper) and 5~GeV$^2$ (lower).
    The effects of TMCs on the (input) scaling distribution
    (dotted curve) are illustrated for the sTMC (dashed) and
    mTMC (double-dot--dashed) prescriptions, and compared
    with the effects on the (input) TD-distribution
    $\xi q^{\rm TD}(\xi)$ (dot-dashed) using the TD approach
    (prescription C, solid).}
\label{fig:F2}
\end{figure}

The effects of TMCs on the $F_2$ structure function are illustrated in
Fig.~\ref{fig:F2} for $Q^2 = 1$ and 5~GeV$^2$ for the various scenarios.
Here the scaling function $x q(x)$ (dotted curve) is used as input in
Eq.~(\ref{eq:F2}) to calculate the target mass corrected function $F_2$.
To translate the results from $\xi$ to $x$, we fix $Q^2$ and extract
the corresponding $x$ for each $(\xi,Q^2)$ pair.
For the sTMC method (prescription A, dashed), in which the upper
limits of the integrals are set to unity, the corrected structure
function becomes smaller at intermediate $x$ values, but larger as
$x \to 1$.
The corrected structure function for the mTMC approach (prescription B,
double-dot--dashed), where the $\xi$ integration is constrained by
$\xi < \xi_0$, display a similar behavior as a function of $x$.
In both cases $F_2$ is clearly nonzero in the $x \to 1$ limit.
The effects are sizable at $Q^2 = 1$~GeV$^2$, but considerably smaller
at $Q^2 = 5$~GeV$^2$, where the differences between the scaling and
target mass corrected functions are more strongly suppressed.
At lower $Q^2$ values, $Q^2 \sim 0.5$~GeV$^2$ (not shown), the
differences between the sTMC and mTMC prescriptions are even more
pronounced, so that here it is important to take into account the
correct kinematics, especially at large $x$.
Note that the sTMC and mTMC curves in Fig.~\ref{fig:F2} were normalized
such that the quark number at finite $Q^2$ is equal to the quark number
at $Q^2 \to \infty$.

The effect of the threshold dependent prescription C is dramatically
different from the other prescriptions.
Specifically, the TD input distribution $q^{\rm TD}$ (dot-dashed curve
in Fig.~\ref{fig:F2}) produces a target mass corrected leading twist
structure function which is exactly zero at $x=1$ ($\xi = \xi_0$)
for all $Q^2$, as required physically.
Another important difference between the sTMC and mTMC approaches,
and the TD prescription, is at intermediate $x$.
Here the latter produces a corrected structure function which is
{\em smaller} than the input, in contrast with the sTMC/mTMC methods,
where the corrected $F_2$ is larger than the input scaling function.
Such differences may be very relevant in phenomenological determinations
of the $x$ dependence of parton distributions at low $Q^2$.

How does the TD prescription C help in the practical extraction of
leading twist parton distributions?
If we define:
\begin{equation}
A_n \equiv \xi_0^{-(n + 3/2)} A_n(Q^2)\ ,
\end{equation}
then since at high $Q^2$ the moments of the quark distributions become
$Q^2$ independent, so too are the Nachtmann moments of the structure
functions, as the multiplicative factor is $\xi$ independent.
Other prescriptions would render a different $\xi_0$, but the important
fact is that once $\mu_2^n(Q^2) = A_n(Q^2)$ for any $Q^2$, this equality
may be set to its value at $Q^2 \to \infty$:
\begin{equation}
\frac{\mu_2^n({\rm finite} \; Q^2)}{A_n({\rm finite} \; Q^2)}
= \frac{\mu_2^n(Q^2 \to \infty)}{A_n(Q^2 \to \infty)}
= \frac{M_2^n}{A_n}\ .
\end{equation}
This makes the approach in prescription C much more useful for
extracting quark distributions from structure function data at
low $Q^2$.

\begin{figure}[htb]
\epsfig{file=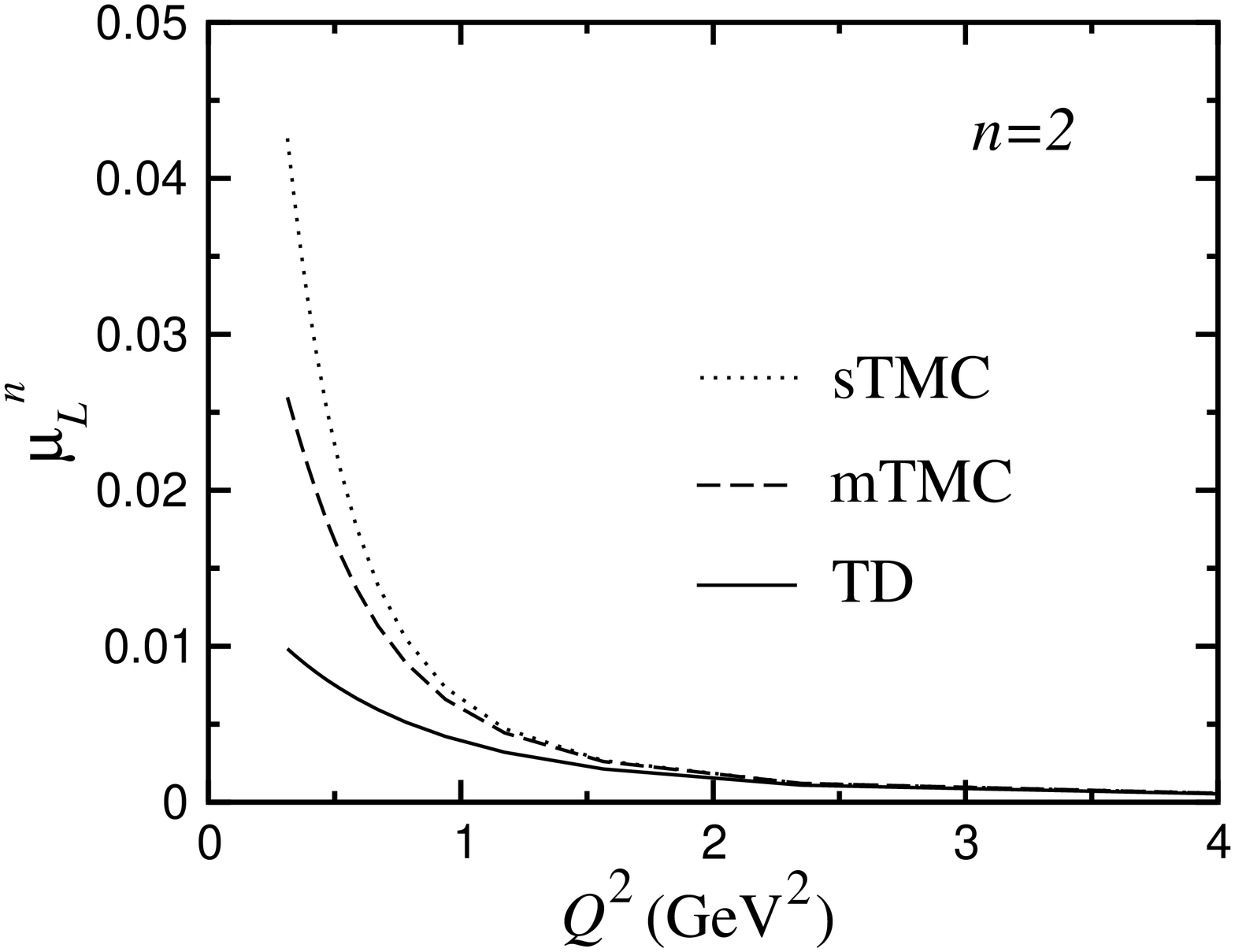,height=12cm,angle=-90}
\epsfig{file=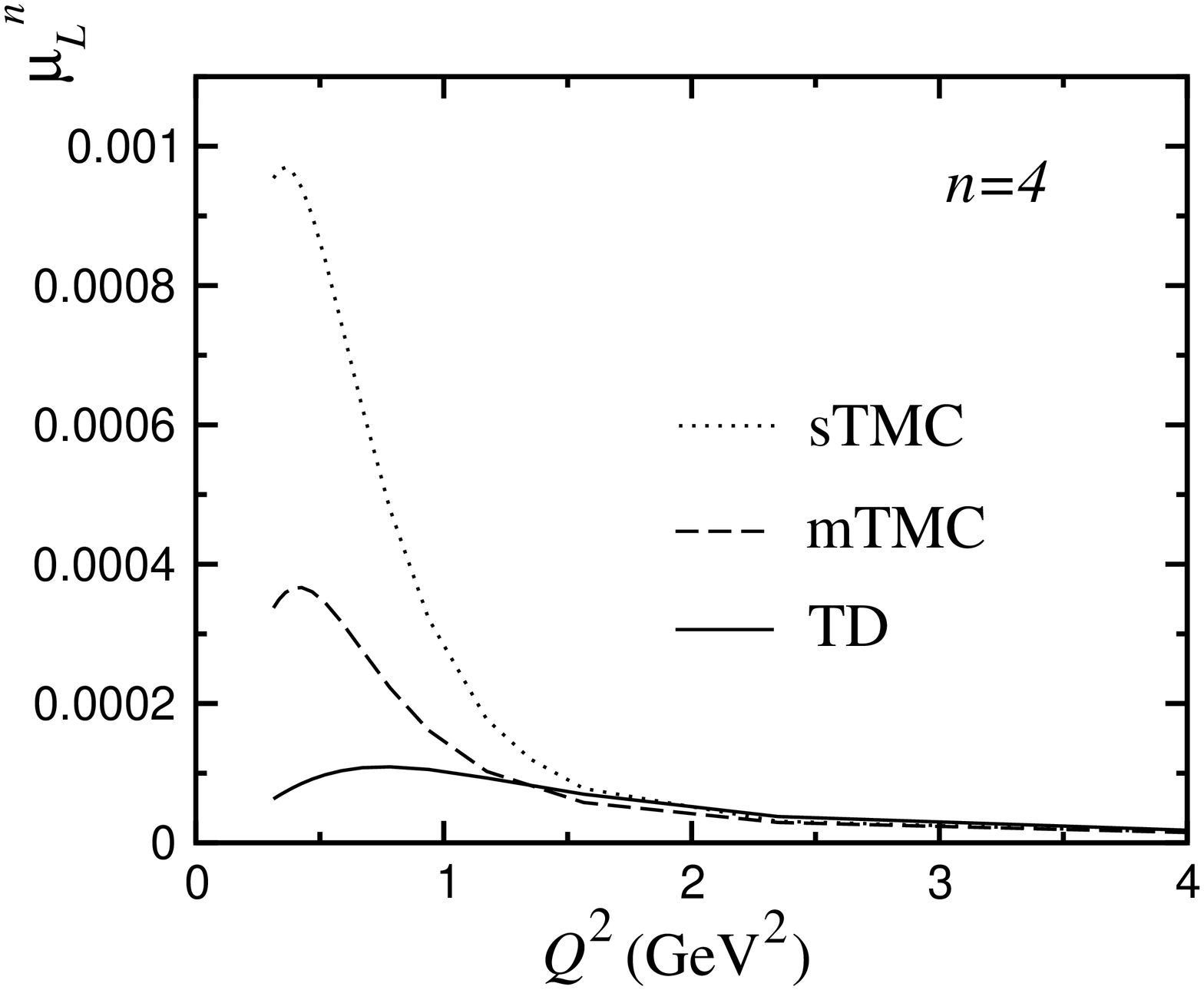,height=12cm,angle=-90}
\caption{Nachtmann moments of the longitudinal structure function
    for $n=2$ (upper) and $n=4$ (lower) as a function of $Q^2$.}
\label{fig:muL}
\end{figure}

A comparison of the different TMC prescriptions for the longitudinal
structure function moments is shown in Fig.~\ref{fig:muL}, for the
$n=2$ and $n=4$ Nachtmann moments.
Here we plot the absolute value of the moments rather than a ratio,
since the moments in the scaling limit are identically zero.
For the sTMC prescription, the corrected moments are strongly $Q^2$
dependent for $Q^2 \lesssim 1$~GeV$^2$ and rise rapidly as $Q^2$
decreases.
The $Q^2$ dependence of the moments for the mTMC approach, with the
upper integration limit being $\xi_0$ rather than unity, is somewhat
weaker, but still quite strong at low $Q^2$.
On the other hand, the moments for the TD prescription C display
significantly smaller $Q^2$ dependence at low $Q^2$.

\begin{figure}[htb]
\epsfig{file=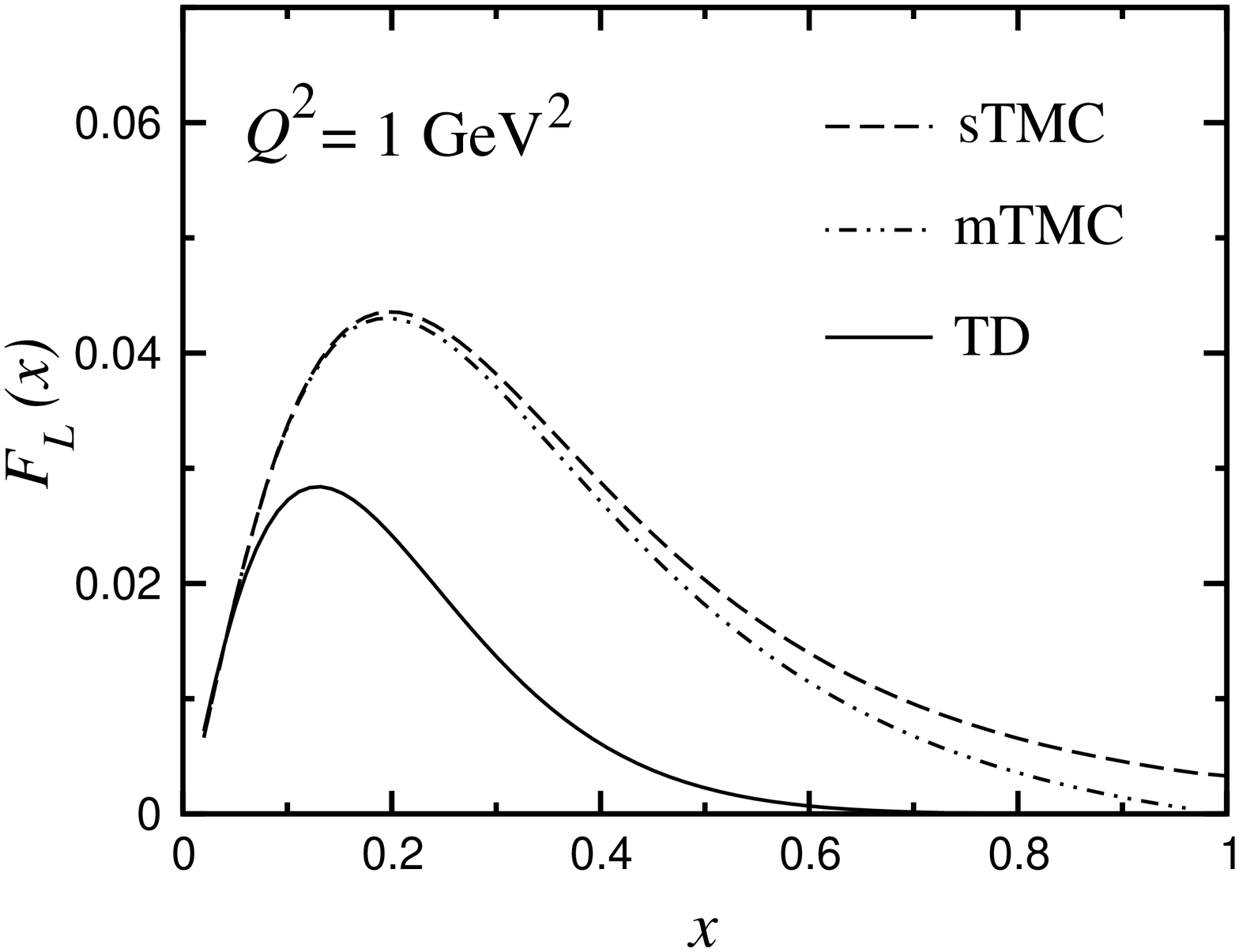,height=12cm,angle=270}
\caption{Longitudinal structure function $F_L$ at $Q^2=1$~GeV$^2$
    for the sTMC (dashed), mTMC (double-dot--dashed) and
    TD-distribution (solid) prescriptions.
    Note that the scaling longitudinal distribution is zero.}
\label{fig:FL}
\end{figure}

Finally, for completion, we show in Fig.~\ref{fig:FL} the $F_L$
structure function at $Q^2 = 1$~GeV$^2$ with TMCs applied according
to the three prescriptions above.
Note that in the scaling limit, the $F_L$ structure function is
identically zero.
For the sTMC and mTMC prescriptions, the corrected structure function
is significantly larger in magnitude than for the TD prescription at
intermediate and large $x$.
For the sTMC case in particular, it is also seen to approach a
nonzero value in the $x \to 1$ limit.
This result suggests that the evaluation of the twist-two part of the
longitudinal structure function at low $Q^2$ may also need to be
reassessed in phenomenological analyses, especially at intermediate
and large $x$.

\section{Conclusion}

In this work we have revisited the long-standing problem of target
mass corrections to nucleon structure functions.
The standard procedure for implementing target mass effects suffers
from the well known threshold problem, in which the corrected, leading
twist structure function does not vanish at $x=1$.
We have proposed a solution to this problem by introducing a
finite-$Q^2$, ``threshold dependent'' parton distribution function
that explicitly depends on the kinematical threshold $\xi_0$,
which is smooth in the entire physical region, and approaches the
ordinary, $Q^2$-independent parton distribution in the limit
$Q^2 \to \infty$.
Our prescription avoids any discontinuities in the parton
distributions and structure functions at finite $Q^2$, and produces
vanishing structure functions as $x \to 1$.
This is true for both the $F_2$ and $F_L$ structure functions.

The Nachtmann moments $\mu_2^n$ of the $F_2$ structure function,
calculated with the threshold dependent distributions $q^{\rm TD}$,
agree with the moments $A_n$ of $q^{\rm TD}$ to within 1\% for the
$n=2, 4$ and 6 moments for $Q^2$ as low as 1~GeV$^2$ and even lower.
In contrast, the deviation for the standard or modified TMC procedure
(sTMC or mTMC prescriptions) is more than an order of magnitude larger
at the same $Q^2$ values, and grows rapidly with increasing $n$.
Furthermore, for $Q^2 > M^2$ one can show analytically that,
at least to ${\cal O}(1/Q^6)$, the moments $\mu_2^n$ and $A_n$
are identical.
Similarly, for the longitudinal structure function $F_L$, the
Nachtmann moments $\mu_L^n$ with the threshold dependent
distribution are considerably smaller ({\em i.e.} closer to the
asymptotic value of zero) than the moments in the sTMC or mTMC
prescriptions.

A consequence of our formulation is that the moments of the
threshold dependent distributions will in general be $M^2/Q^2$
dependent.
This dependence is not associated with either perturbative QCD
effects or higher twists, but comes entirely from the leading twist,
target mass effects.
Our analysis suggests that it may be necessary to reassess the
interpretation of a parton distribution in the presence of the
finite $M^2/Q^2$, or $\xi$, corrections, as well as the implementation
of the $q^{\rm TD}$ distributions in the $Q^2$ evolution equations.
We will address these problems in future work \cite{future}.
At the same time, our numerical results give impetus to investigating
the impact of TMCs on phenomenological fits to structure functions
at low $Q^2$ \cite{NewFits} and the extraction of twist-two parton
distributions.

\acknowledgments

We would like to thank J.~Bl\"umlein, M.~E.~Christy, C.~E.~Keppel,
and A.~W.~Thomas for helpful discussions.
This work was supported by the U.S. Department of Energy contract
DE-AC05-84ER40150, under which the Southeastern Universities
Research Association (SURA) operates the Thomas Jefferson National
Accelerator Facility (Jefferson Lab). FMS is also supported by
FAPESP (03/10754-0) and CNPq (308932/2003-0).



\begin{thebibliography}{99}

\bibitem{PRep}
W.~Melnitchouk, R.~Ent and C.~Keppel,
Phys.\ Rept.\  {\bf 406}, 127 (2005).

\bibitem{HTanal}
Z.-E.~Meziani, W.~Melnitchouk, J.-P.~Chen, S.~Choi {\em et al.},
Phys.\ Lett.\ B {\bf 613}, 148 (2005);
%
M.~Osipenko {\em et al.}, Phys.\ Lett.\ B {\bf 609}, 259 (2005);
%
A.~Deur {\em et al.}, Phys.\ Rev.\ Lett.\ {\bf 93}, 212001 (2004);
%
M.~Osipenko, S.~Simula, W.~Melnitchouk {\em et al.}, Phys.\ Rev.\
D {\bf 71}, 054007 (2005);
%
S.I.~Alekhin, S.A.~Kulagin and S.~Liuti, Phys.\ Rev.\ D {\bf 69},
114009 (2004);
%
N.~Bianchi, A.~Fantoni and S.~Liuti, Phys.\ Rev.\ D {\bf 69},
014505 (2004);
%
M.~Osipenko, W.~Melnitchouk, S.~Simula, S.~Kulagin and G.~Ricco,
Nucl.\ Phys.\ A {\bf 766}, 142 (2006).

\bibitem{GP76}
H.~Georgi and H.~D.~Politzer,
Phys.\ Rev.\ D {\bf 14}, 1829 (1976).

\bibitem{Bluemlein}
J.~Bl\"umlein and A.~Tkabladze,
Nucl.\ Phys.\ B {\bf 553}, 427 (1999);
%
J.\ Phys.\ G {\bf 25}, 1553 (1999).

\bibitem{Nac73}
O.~Nachtmann,
Nucl.\ Phys.\ B {\bf 63}, 237 (1973).

\bibitem{Ell76}
R.~K.~Ellis, R.~Petronzio and G.~Parisi,
Phys.\ Lett.\ B {\bf 64}, 97 (1976);
%
R.~Barbieri, J.~R.~Ellis, M.~K.~Gaillard and G.~G.~Ross,
Phys.\ Lett.\ B {\bf 64}, 171 (1976);
Nucl.\ Phys.\ B {\bf 117}, 50 (1976).

\bibitem{GTW77}
D.~J.~Gross, S.~B.~Treiman and F.~A.~Wilczek,
Phys.\ Rev.\ D {\bf 15}, 2486 (1977).

\bibitem{BJT79}
K.~Bitar, P.~W.~Johnson and W.~k.~Tung,
Phys.\ Lett.\ B {\bf 83}, 114 (1979).

\bibitem{JT80}
P.~W.~Johnson and W.~K.~Tung,
Print-79-1018 (Illinois Tech) {\it Contribution to Neutrino '79, Bergen,
Norway, June 18-22, 1979}.

\bibitem{Fra80}
W.~R.~Frazer and J.~F.~Gunion,
Phys.\ Rev.\ Lett.\  {\bf 45}, 1138 (1980).

\bibitem{DGP77}
A.~De Rujula, H.~Georgi and H.~D.~Politzer,
Phys.\ Rev.\ D {\bf 15}, 2495 (1977).

\bibitem{F2JL}
I.~Niculescu {\it et al.},
Phys.\ Rev.\ Lett.\  {\bf 85}, 1182, 1186 (2000).

\bibitem{Kretzer}
S.~Kretzer and M.~H.~Reno,
Phys.\ Rev.\ D {\bf 69}, 034002 (2004).

\bibitem{Detmold}
W.~Detmold,
Phys.\ Lett.\ B {\bf 632}, 261 (2006).

\bibitem{future}
F.~M.~Steffens {\em et al.}, in preparation.

\bibitem{NewFits}
M.~E.~Christy {\em et al.}, private communication.

\end{thebibliography}
\end{document}